\documentclass[prl,twocolumn,superscriptaddress,showpacs,amssymb,amsmath,amsfonts,aps]{revtex4}
\usepackage{graphicx}
\usepackage{dcolumn}
\begin{document}
\title{$Q^2$ Dependence of Quadrupole Strength in the $\gamma^*p\rightarrow\Delta^+(1232)\rightarrow p \pi^0$ Transition}

\newcommand*{\mpaa}{|M_{1+}|^2}
\newcommand*{\mpbb}{|E_{1+}|^2}
\newcommand*{\mpcc}{|S_{1+}|^2}
\newcommand*{\mpdd}{|M_{1-}|^2}
\newcommand*{\mpee}{|E_{0+}|^2}
\newcommand*{\mpff}{|S_{0+}|^2}
\newcommand*{\mpgg}{|S_{1-}|^2}
\newcommand*{\mpa}{M_{1+}}
\newcommand*{\mpb}{E_{1+}}
\newcommand*{\mpc}{S_{1+}}
\newcommand*{\mpd}{M_{1-}}
\newcommand*{\mpe}{E_{0+}}
\newcommand*{\mpf}{S_{0+}}
\newcommand*{\mpg}{S_{1-}}

\newcommand*{\uvch}{University of Virginia, Department of Physics, Charlottesville, VA 22903, USA}
\newcommand*{\cnuva}{Christopher Newport University, Newport News, VA 23606, USA}
\newcommand*{\jlab}{Thomas Jefferson National Accelerator Facility, 12000 Jefferson Avenue, Newport News, VA 23606, USA}
\newcommand*{\yerevan}{Yerevan Physics Institute, 375036 Yerevan, Armenia}
\newcommand*{\asuaz}{Arizona State University, Department of Physics and Astronomy, Tempe, AZ 85287, USA}
\newcommand*{\cmupa}{Carnegie Mellon University, Department of Physics, Pittsburgh, PA 15213, USA}
\newcommand*{\cuawdc}{Catholic University of America, Department of Physics, Washington D.C., 20064, USA}
\newcommand*{\cwm}{College of William and Mary, Department of Physics, Williamsburg, VA 23187, USA}
\newcommand*{\duke}{Duke University, Physics Bldg. TUNL, Durham, NC27706, USA}
\newcommand*{\edinburgh}{Department of Physics and Astronomy, Edinburgh University, Edinburgh EH9 3JZ, United Kingdom}
\newcommand*{\fiu}{Florida International University, Miami, FL 33199, USA}
\newcommand*{\fsu}{Florida State University, Department of Physics, Tallahassee, FL 32306, USA}
\newcommand*{\gwudc}{George Washington University, Department of Physics, Washington D. C., 20052 USA}
\newcommand*{\frascati}{Istituto Nazionale di Fisica Nucleare, Laboratori Nazionali di Frascati, P.O. 13, 00044 Frascati, Italy}
\newcommand*{\genova}{Istituto Nazionale di Fisica Nucleare, Sezione di Genova
 e Dipartimento di Fisica dell'Universita, 16146 Genova, Italy}
\newcommand*{\itep}{Institute of Theoretical and Experimental Physics, 25 B. Cheremushkinskaya, Moscow, 117259, Russia}
\newcommand*{\ipn}{Institut de Physique Nucleaire d'Orsay, IN2P3, BP 1, 91406 Orsay, France}
\newcommand*{\jmuva}{James Madison University, Department of Physics, Harrisonburg, VA 22807, USA}
\newcommand*{\knukorea}{Kyungpook National University, Department of Physics, Taegu 702-701, South Korea}
\newcommand*{\mmit}{M.I.T.-Bates Linear Accelerator, Middleton, MA 01949, USA}
\newcommand*{\nsuva}{Norfolk State University, Norfolk VA 23504, USA}
\newcommand*{\ohio}{Ohio University, Department of Physics, Athens, OH 45701, USA}
\newcommand*{\oduva}{Old Dominion University, Department of Physics, Norfolk VA 23529, USA}
\newcommand*{\rpi}{Rensselaer Polytechnic Institute, Department of Physics, Troy, NY 12181, USA}
\newcommand*{\rubltx}{Rice University, Bonner Lab, Box 1892, Houston, TX 77251, USA}
\newcommand*{\sphn}{CEA Saclay, DAPNIA-SPhN, F91191 Gif-sur-Yvette Cedex, France}
\newcommand*{\ucla}{University of California at Los Angeles, Department of Physics and Astonomy, Los Angeles, CA 90095-1547, USA}
\newcommand*{\connecticut}{University of Connecticut, Physics Department, Storrs, CT 06269, USA}
\newcommand*{\umma}{University of Massachusetts, Department of Physics, Amherst, MA 01003, USA}
\newcommand*{\unhdurham}{University of New Hampshire, Department of Physics, Durham, NH 03824, USA}
\newcommand*{\uppa}{University of Pittsburgh, Department of Physics and Astronomy, Pittsburgh, PA 15260, USA}
\newcommand*{\urva}{University of Richmond, Department of Physics, Richmond, VA 23173, USA}
\newcommand*{\usc}{University of South Carolina, Department of Physics, Columbia, SC 29208, USA}
\newcommand*{\utep}{University of Texas at El Paso, Department of Physics, El Paso, Texas 79968, USA}
\newcommand*{\vpsu}{Virginia Polytechnic and State University, Department of Physics, Blacksburg, VA 24061, USA}

\affiliation{\uvch}
\affiliation{\cnuva}
\affiliation{\jlab}
\affiliation{\yerevan}
\affiliation{\asuaz}
\affiliation{\cmupa}
\affiliation{\cuawdc}
\affiliation{\cwm}
\affiliation{\duke}
\affiliation{\edinburgh}
\affiliation{\fiu}
\affiliation{\fsu}
\affiliation{\gwudc}
\affiliation{\frascati}
\affiliation{\genova}
\affiliation{\itep}
\affiliation{\ipn}
\affiliation{\jmuva}
\affiliation{\knukorea}
\affiliation{\mmit}
\affiliation{\nsuva}
\affiliation{\ohio}
\affiliation{\oduva}
\affiliation{\rpi}
\affiliation{\rubltx}
\affiliation{\sphn}
\affiliation{\ucla}
\affiliation{\connecticut}
\affiliation{\umma}
\affiliation{\unhdurham}
\affiliation{\uppa}
\affiliation{\urva}
\affiliation{\usc}
\affiliation{\utep}
\affiliation{\vpsu}

\author{K.~Joo}\affiliation{\uvch}
\author{L.C.~Smith}\affiliation{\uvch}
\author{V.D.~Burkert}\affiliation{\jlab}
\author{R.~Minehart}\affiliation{\uvch}
\author{I.G.~Aznauryan}\affiliation{\yerevan}
\author{L.~Elouadrhiri}\affiliation{\cnuva}\affiliation{\jlab}
\author{S.~Stepanyan}\affiliation{\yerevan}\affiliation{\oduva}
\author{G.S.~Adams}\affiliation{\rpi}
\author{M.J.~Amaryan}\affiliation{\yerevan}
\author{E.~Anciant}\affiliation{\sphn}
% ...
\collaboration{The CLAS Collaboration}
\noaffiliation
\begin{abstract}
{Models of baryon structure predict a small quadrupole deformation of the 
nucleon due to residual tensor forces between quarks or distortions from 
the pion cloud.  Sensitivity to quark versus pion degrees of freedom occurs through
the $Q^2$ dependence of the magnetic ($M_{1+}$), electric ($E_{1+}$), and 
scalar ($S_{1+}$) multipoles in the $\gamma^* p \rightarrow \Delta^+ \rightarrow p \pi^0$ transition.
We report new experimental values for the ratios $E_{1+}/M_{1+}$ and 
$S_{1+}/M_{1+}$ over the range $Q^2$=\,0.4-1.8~GeV$^2$, extracted
from precision $p(e,e\,'p)\pi^{\circ}$ data using a truncated multipole expansion.  
Results are best described by recent unitary models in which the pion cloud plays a dominant role.}
\end{abstract}

\pacs{PACS : 13.60.Le, 13.40.Gp, 14.20.Gk}
\maketitle
Electroproduction of nucleon resonances provides unique information
about the internal dynamics of baryons. 
For the $\gamma^* N\rightarrow\Delta(1232)\rightarrow N\pi$ transition, a long-standing problem is to 
achieve a consistent experimental and theoretical description of the electric and scalar quadrupole 
multipoles $E_{1+}$ and $S_{1+}$, and the magnetic dipole $M_{1+}$.  
Within $SU(6)$ models this transition is mediated by a single quark spin flip in the $L=0$
nucleon ground state, leading to $M_{1+}$ dominance and $E_{1+} = S_{1+} \equiv 0$. 
QCD-motivated constituent quark models introduce a tensor force from the inter-quark hyperfine 
interaction, which leads to a $d$-state admixture in the baryon wave
function \cite{isg92}. As a result small but non-zero values for $E_{1+}$ and $S_{1+}$
are predicted \cite{isg92,cap90}. Stronger contributions are expected from the pion 
cloud \cite{ber88,wal97,sil00,amo00} 
or from two-body exchange currents \cite{buc98}.
Finally, quark helicity conservation in pQCD requires $E_{1+} = M_{1+}$ as $Q^2 \rightarrow \infty$. 

Determination of the ratios $R_{EM}=E_{1+}/M_{1+}$ 
and $R_{SM}=S_{1+}/M_{1+}$ has been the aim of a considerable number of 
experiments in the past. While theoretical models have become more refined, most 
previous measurements have large systematic and statistical errors or 
significantly limited kinematic coverage. 
A new program using the CEBAF Large Acceptance Spectrometer (CLAS) \cite{bro00} at 
Jefferson Lab has been inaugurated to vastly improve the systematic 
and statistical precision by covering 
a wide kinematic range of four-momentum transfer $Q^2$ and invariant 
mass $W$, and by subtending the full angular range of the resonance decay 
into the $\pi N$ final state. 

This Letter reports the first CLAS results for $R_{EM}$ and $R_{SM}$ obtained 
from a partial wave analysis of the $p(e,e'p)\pi^0$ reaction for 
$Q^2$=0.4-1.8~GeV$^2$. This $Q^2$ range
explores distance scales where dynamical breaking of chiral symmetry may introduce
collective degrees of freedom in the nucleon.  Interest
in chiral models recently increased after photo-pion measurements from LEGS \cite{bla00} and MAMI \cite{bec01} 
found $R_{EM}=-3.1\%$ and -2.5\%, respectively at $Q^2=0$, which is substantially larger than
constituent quark model predictions \cite{isg92,cap90}.  Chiral bag \cite{ber88} and soliton 
models \cite{wal97,sil00,amo00} 
in which quark confinement occurs through non-linear interactions with the pion cloud, 
generally find $R_{EM}$ in the range -1\% to -5\% at $Q^2=0$. 
Chiral effective field theories \cite{gel99} and unitary \cite{dre99} 
and dynamical reaction models \cite{sat01,kam01} that employ pion rescattering at the 
$\gamma^* N\Delta$ vertex, predict meson degrees of freedom should 
enhance the quadrupole strength at low  $Q^2$ and strongly affect 
the $Q^2$ dependence of $R_{EM}$ and $R_{SM}$.  

Under the one-photon-exchange approximation, the pion electroproduction cross 
section factorizes as follows:
\begin{equation}
\frac{d\,^5\sigma}{dE_{e'} d\Omega_{e'} d\Omega^*_{\pi}} = \Gamma_v\,\frac{d\,^2\sigma}{d\Omega^*_{\pi}},
\label{eq:1}
\end{equation}
where $\Gamma_v$ is the virtual photon flux. For an unpolarized beam and target the 
center-of-mass (cm) differential cross section $d\,^2\sigma/d\Omega^*_{\pi}$ depends 
on the transverse $\epsilon$ and longitudinal $\epsilon_L$  polarization of the virtual photon 
through four structure functions: $\sigma_T, \sigma_L$, and the interference 
terms $\sigma_{LT}$ and $\sigma_{TT}$:
\begin{eqnarray}
\frac{d\,^2\sigma}{d\Omega^*_{\pi}}&=&\frac{p^*_{\pi}}{k_{\gamma}^*}(\sigma_T+\epsilon_L\sigma_L+\epsilon\,\sigma_{TT}\,\sin^2\theta^*_{\pi}\,\cos\,2\phi^*_{\pi} \nonumber \\
&+&\sqrt{2\epsilon_L(\epsilon+1)}\,\sigma_{LT}\,\sin\,\theta^*_{\pi}\,\cos\,\phi^*_{\pi}),
\label{eq:2}
\end{eqnarray}
where ($p_{\pi}^*,\theta^*_{\pi},\phi^*_{\pi}$) are the $\pi^0$ cm momentum, polar, and azimuthal angles,
$\epsilon_L=(Q^2/|k^*|^2)\epsilon$, and $|k^*|$ and $k_{\gamma}^*$ are the virtual photon cm momentum 
and real photon cm equivalent energy. A partial wave expansion of the structure functions 
using Legendre polynomials $P_l(\cos\theta^*_{\pi})$ gives (for $\textit{s}$- and $\textit{p}$-waves):
\begin{subequations}
\label{eq3}
\begin{eqnarray}
\sigma_T+\epsilon_L\sigma_L &=& A_{0}+A_{1}\,P_1+A_{2}\,P_2 \label{eq:3a}\\
\sigma_{TT} &=& C_0 \label{eq:3b}\\
\sigma_{LT} &=& D_0+D_1\,P_1. \label{eq:3c}
\end{eqnarray}
\end{subequations}
The weak quadrupole $\mpb$ and $\mpc$ transitions are accessible only through their interference 
with the dominant $\mpa$.  To simplify the analysis, a truncated multipole expansion is used,
in which only terms involving $M_{1+}$ are 
retained.  Thus, $\mpaa$ and its projection onto the other $\textit{s}$- and $\textit{p}$-wave multipoles 
$\mpb,\mpc,\mpd,\mpe,\mpf$ are given in terms of the six partial-wave coefficients by~ \cite{ras89}:
\begin{subequations}
\label{eq4}
\begin{eqnarray}
\mpaa & = & A_0/2 \label{eq:4a}\\
Re(\mpb\,\mpa^*) &=& (A_2-2\,C_0/3)/8 \label{eq:4b} \\
Re(\mpd\,\mpa^*) &=& -(A_2+2\,(A_0+C_0))/8 \label{eq:4c}\\
Re(\mpe\,\mpa^*) &=& A_1/2 \label{eq:4d}\\
Re(\mpf\,\mpa^*) &=& D_0 \label{eq:4e} \\
Re(\mpc\,\mpa^*) &=& D_1/6. \label{eq:4f}
\end{eqnarray}
\end{subequations}
In accordance with previous analyses \cite{bec01,fro99} we define $R_{EM}$ and $R_{SM}$ as:
\begin{eqnarray}
R_{EM}&=&Re(\mpb\,\mpa^*)/\mpaa  \\
R_{SM}&=&Re(\mpc\,\mpa^*)/\mpaa.
\end{eqnarray}
Near the $\Delta(1232)$ mass, where the isospin 3/2 channel dominates and 
$Re(M^{(3/2)}_{1+})$ vanishes, $R_{EM}\approx~Im(E^{(3/2)}_{1+})~/~Im(M^{(3/2)}_{1+})$
and similarly for $R_{SM}$.  The contribution to $R_{EM}$ from $Re(M^{(1/2)}_{1+})$ 
was estimated in \cite{bec01} to be $< 0.5\%$ absolute at $Q^2=0$, and is $<0.3\%$ for
$Q^2<2.0~$GeV$^2$ \cite{dre99}.
 
\begin{figure}[h]
\includegraphics[scale=0.45]{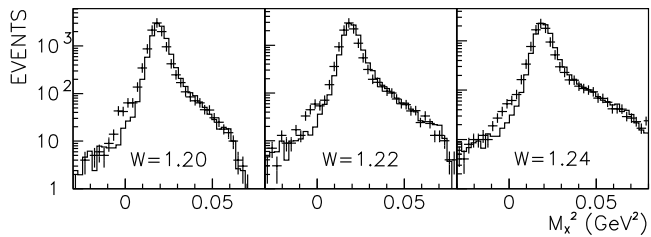}
\caption{Experimental $p(e,e'p)X$ missing mass for invariant mass $W$ bins around
the $\Delta(1232)$ (note logarithmic scale). Solid line: Simulation of CLAS response 
to $p(e,e'p)\pi^o$ reaction. The small shoulder at $M_x^2=0$ is due to 
residual $ep\rightarrow e'p\gamma$ events which survive the kinematic cuts.}
\label{fig1}
\end{figure}

The present measurement used two beam energies (1.645 and 2.445 GeV) to
cover the interval $Q^2$=0.4-1.8~GeV$^2$. A 2.5~nA beam was delivered onto a 
4.0~cm long liquid-hydrogen target at 100\% duty factor. 
Particles were detected in CLAS, which provides momentum coverage down to 
0.1~GeV/c over a polar angle ($\theta$) range $8^{\circ}-142^{\circ}$ and covers 
nearly 80\% of the azimuthal angle $\phi$.  A hardware electron trigger was 
formed using threshold \v{C}erenkov counters and electromagnetic calorimeters, 
while protons were identified using time-of-flight.  Software fiducial cuts 
excluded regions of non-uniform detector response, and the acceptance and 
tracking efficiency were determined using a Monte-Carlo simulation and a 
GEANT model of the detector.  After kinematic corrections the invariant mass 
$W$ resolution was $\sigma_W\approx$~8-10 MeV. 

Coincident protons were used to identify the $\pi^0$.  A typical missing mass spectrum is 
compared in Fig.~1 to a GEANT simulation that incorporates radiation 
effects and detector resolution, using a phenomenological model of the $p(e,e'p)\pi^0$ reaction.
Good agreement with the width and radiative tail of the $\pi^0$ peak is seen.  Background 
from elastic Bethe-Heitler radiation was suppressed using a combination of cuts on missing 
mass and $\phi^*_{\pi}$ near $M_x^2=0$ and $\phi^*_{\pi}=0^{\circ}$.  A cut of 
$-0.01\leq M_x^2(\textrm{GeV}^2)\leq 0.08$ 
was used to select the $p\pi^0$ final state.   Target window backgrounds and proton 
scattering from the torus coils were suppressed with cuts on the reconstructed $e'p$ target vertex.

\begin{figure}[h]
\includegraphics[scale=0.45]{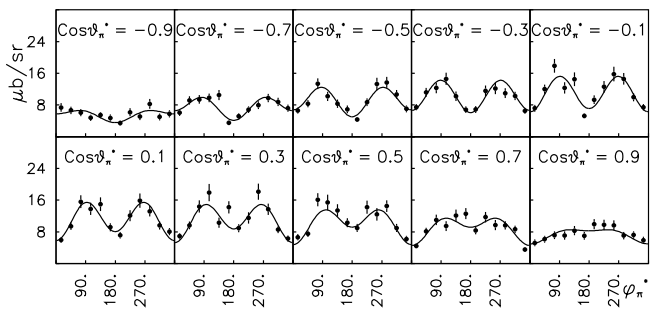}
\caption{Typical $\phi^*_{\pi}$ dependence for the $p(e,e\,'\,p)\pi^0$ cross sections at 
$Q^2$=0.9~GeV$^2$ and $W$=1.22~GeV.  Solid line: Fit to data according to Eq.~(\ref{eq:2}).
Errors are statistical only.}
\label{fig2}
\end{figure}

Systematic errors in the electron kinematics, acceptance and radiative corrections were determined
by measuring inclusive ($e,e^{\prime}$) elastic and inelastic cross sections simultaneously 
with the exclusive data.  The results agreed to within 5\% with parameterizations of previous 
measurements.  Determination of the $\pi^{\circ}p$ cm angles 
($\theta^*_{\pi},\phi^*_{\pi}$) was affected 
by residual $ep\rightarrow e'p\,\gamma$ backgrounds, radiative and kinematic corrections and 
proton multiple scattering.
These systematic effects were estimated by varying cuts on missing mass, 
target vertex reconstruction, and fiducial acceptance. Model dependence of the acceptance
and radiative corrections was studied in detail and included in the systematic error.

\begin{figure}[h]
\includegraphics[scale=0.42]{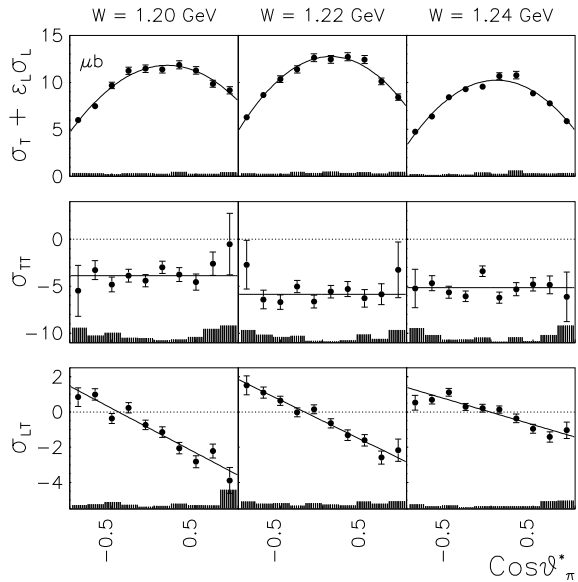}
\caption{Structure functions versus $\cos\,\theta^*_{\pi}$ extracted for 
the $p(e,e\,'\,p)\pi^o$ reaction at $Q^2$= 0.9~GeV$^2$.
Solid line: Legendre polynomial fit to the data using Eq.~(\ref{eq3}). Shaded bars show
systematic errors.}
\label{fig3}
\end{figure}

Typical cross sections obtained after radiative corrections are shown in 
Fig.~\ref{fig2} for $W=1.22$~GeV and illustrates the complete out-of-plane 
$\phi^*_{\pi}$ coverage possible with CLAS.  The presence of
non-zero $\sigma_{TT}$ and $\sigma_{LT}$ strength is clearly indicated by the 
$\cos\,2\phi^*_{\pi}$ and $\cos \phi^*_{\pi}$ modulation of the cross 
sections.  These terms were separated 
from $\sigma_T+\epsilon_L\sigma_L$ by fitting the $\phi^*_{\pi}$ 
distributions with the form in Eq.~(\ref{eq:2}).  The extracted structure functions
are shown in Fig.~\ref{fig3} for several $W$ bins around the $\Delta(1232)$ peak.  
Fits to the $\cos\,\theta^*_{\pi}$ dependence using Eq.~(\ref{eq3}) are 
indicated by the solid curves.  Inclusion of $\textit{d}$-waves, which would
lead to deviations from the linear behavior for $\sigma_{TT}$ and $\sigma_{LT}$ 
in Fig.~\ref{fig3}, did not improve the fit.

\begin{figure}[h]
\includegraphics[scale=0.4]{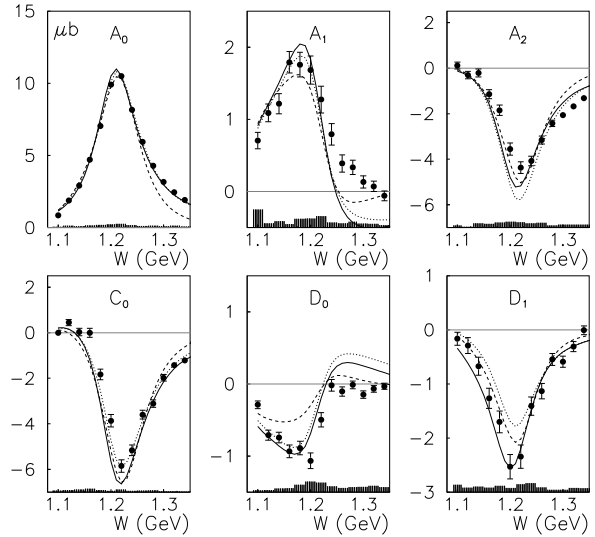}
\caption{$W$ dependence of the Legendre coefficients obtained from structure 
function fits at $Q^2$=0.9~GeV$^2$. The curves show model predictions 
($\textit{s}$- and $\textit{p}$-wave multipoles only) from 
MAID98 \cite{dre99} (dotted), MAID00 \cite{m2000} (solid) and Sato-Lee \cite{sat01} (dashed).
Shaded bars show systematic errors.}
\label{fig4}
\end{figure}

Figure~\ref{fig4} shows the $W$ dependence of the partial wave coefficients obtained from
the structure function fits.  The data are compared to calculations of 
Drechsel\ $\textit{et~al.}$ \cite{dre99} (MAID) and Sato and Lee (SL)  \cite{sat01}. These
models include unitarized contributions from Born diagrams and vector meson exchange, with
the model parameters fitted to previous photo- and electroproduction data. The 
curves show predicted contributions from all $\textit{s}$- and $\textit{p}$-wave multipoles.  
For the $A_0$ coefficient, which is dominated by the well-known 
$|M_{1+}|^2$, both SL and MAID describe the shape and magnitude quite well for $W<1.26~$GeV. 
The neglect of higher-mass resonances in the SL model is clearly evident for 
$W>1.26~$GeV.  $A_1$ and $D_0$ are dominated by the interference between
$M_{1+}$ and the non-resonant electric and scalar $\textit{s}$-wave multipoles 
$E_{0+}$ and $S_{0+}$. Our results are clearly sensitive to differences between
the models, which arise partly from the treatment of backgrounds.

The quadrupole interference terms $Re(\mpb\mpa^*)$ and $Re(\mpc\mpa^*)$ were 
extracted from the $A_2$, $C_0$ and $D_1$ coefficients using Eq.~(\ref{eq:4b}) 
and Eq.~(\ref{eq:4f}), while $\mpaa$ was determined using Eq.~(\ref{eq:4a}).
The ratios $R_{EM}$ and $R_{SM}$ were determined at $W$=1.20, 1.22 and 1.24~GeV and 
averaged to smooth statistical fluctuations.  Errors arising from the $M_{1+}$ 
dominance assumption and the averaging procedure were estimated by fitting 
`pseudo-data' generated from the MAID and SL models and binned identically 
to the CLAS data.  The fitted terms were then compared to those calculated from 
the model input multipoles.  Our typical(worst) absolute truncation error 
(including model dependence) was $0.3(0.7)\%$ for $R_{EM}$ and $0.1(0.5)\%$ 
for $R_{SM}$ over the $Q^2$ range of this experiment, with the error generally
increasing with $Q^2$ due to the larger relative importance of neglected 
non-resonant multipoles.  Results for each $Q^2$ bin are listed
in Table \ref{tab:1}.  Note that measurements at the same $Q^2$ but different beam energies agree
within the uncertainties, lending credence to the accuracy of the corrections.

Figure~5 summarizes the $Q^2$ dependence of the available $R_{EM}$ and $R_{SM}$ data
compared to recent model calculations.  Our results show no $Q^2$ dependence for $R_{EM}$, 
in contrast to the rapid falloff to zero predicted 
by chiral-quark/soliton models ($\chi$QSM)  \cite{sil00,amo00}.  Although 
motivated by chiral symmetry, these models ignore the $\Delta\rightarrow\pi N$ decay
and rescattering effects.  The two relativistic quark model $R_{EM}$ curves, 
RQM1 \cite{war90} and RQM2 \cite{azn93}, agree at $Q^2=0$, but strongly diverge for $Q^2>0$, while 
the zero crossing seen 
in  \cite{azn93} is excluded by the CLAS data.  Our overall $R_{EM}\approx -2\%$ is
consistent with recent measurements both at lower $Q^2$ \cite{bla00,bec01,mer01}, and at
higher $Q^2$ \cite{fro99}.  The Coulomb 
quadrupole ratio $R_{SM}$ is significantly larger in magnitude and shows
a strong $Q^2$ dependence.  While the chiral models and RQM2 do somewhat better 
in comparison with $R_{SM}$, so far no quark or chiral soliton model is able to
successfully describe both $R_{EM}$ and $R_{SM}$.

\begingroup
\squeezetable
\begin{table}
\caption{\label{table1}Quadrupole/magnetic dipole ratios for the $\gamma^* N\rightarrow \Delta(1232)$ 
transition from partial wave fits at invariant momentum transfer $Q^2$ and beam 
energy $E_e$.  The first error is statistical, while the experiment-related systematic 
effects are included in the second error. \label{tab:1}}
\begin{ruledtabular} 
\begin{tabular}{cccc}
$Q^2$&$E_e$&$Re(E_{1+}/M_{1+}]$&$Re(S_{1+}/M_{1+})$ \\ 
(GeV$^2$)&(GeV)&(\%)&(\%)\\ \cline{1-4}
0.40 & 1.645 &-3.4 $\pm$ 0.4 $\pm$ 0.4 & -5.6 $\pm$ 0.4 $\pm$ 0.6 \\ 
0.52 & 1.645 &-1.6 $\pm$ 0.4 $\pm$ 0.4 & -6.4 $\pm$ 0.4 $\pm$ 0.5 \\  
0.65 & 1.645 &-1.9 $\pm$ 0.5 $\pm$ 0.5 & -6.9 $\pm$ 0.6 $\pm$ 0.5 \\ 
0.75 & 1.645 &-2.1 $\pm$ 0.6 $\pm$ 0.7 & -7.4 $\pm$ 0.8 $\pm$ 0.5 \\   
0.90 & 1.645 &-1.8 $\pm$ 0.6 $\pm$ 0.4 & -8.4 $\pm$ 0.9 $\pm$ 0.4 \\ 
0.65 & 2.445 &-2.0 $\pm$ 0.4 $\pm$ 0.4 & -6.6 $\pm$ 0.4 $\pm$ 0.2 \\  
0.75 & 2.445 &-1.6 $\pm$ 0.5 $\pm$ 0.5 & -6.0 $\pm$ 0.4 $\pm$ 0.2 \\   
0.90 & 2.445 &-1.8 $\pm$ 0.4 $\pm$ 0.3 & -7.2 $\pm$ 0.4 $\pm$ 0.1 \\  
1.15 & 2.445 &-1.6 $\pm$ 0.5 $\pm$ 0.3 & -7.9 $\pm$ 0.5 $\pm$ 0.4 \\  
1.45 & 2.445 &-2.4 $\pm$ 0.7 $\pm$ 0.4 & -7.7 $\pm$ 0.9 $\pm$ 0.7 \\  
1.80 & 2.445 &-0.9 $\pm$ 1.1 $\pm$ 0.7 &-11.6 $\pm$ 1.6 $\pm$ 1.5 \\ 
\end{tabular}
\end{ruledtabular}
\end{table}
\endgroup
\begin{figure}[h]
\includegraphics[scale=0.46]{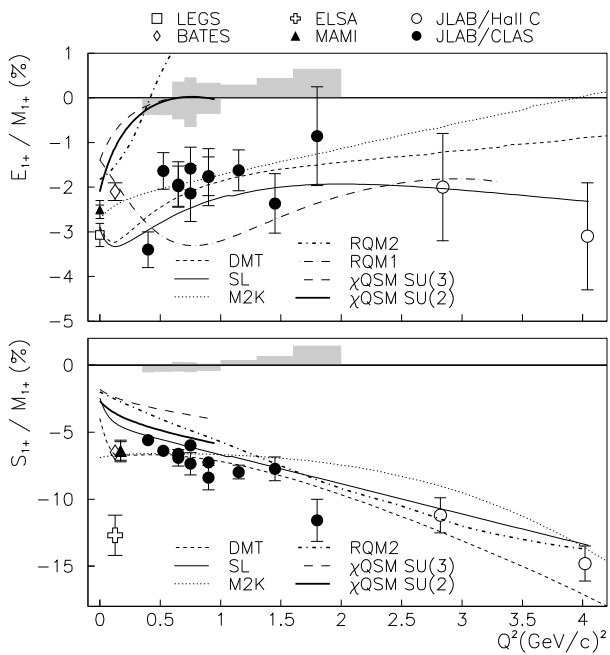}
\caption{$Q^2$ dependence of the electric ($E_{1+}$) and scalar ($S_{1+}$) 
quadrupole/magnetic dipole ratios from this experiment ($\bullet$).  Shaded bands show systematic
errors for the two beam energies listed in Table I. Truncation/averaging errors are discussed in the
text.  Other points are from BATES \cite{mer01}, ELSA \cite{kal97}, 
JLAB/Hall C \cite{fro99}, LEGS \cite{bla00} and MAMI \cite{pos01,bec01}.
The curves show recent model calculations (see text): $\chi$QSM \cite{sil00}, DMT \cite{kam01},
SL \cite{sat01}, M2K \cite{m2000,kam01}, RQM1 \cite{war90}, RQM2 \cite{azn93}.}
\label{fig5}
\end{figure}

Dynamical pion rescattering models calculate a meson `dressed' $\gamma^* N\Delta$ vertex in terms
of the underlying `bare' photocoupling form factors. Sato and Lee  \cite{sat01} fitted their dynamical 
model to photo-pion observables \cite{bec01} and the JLAB/Hall C cross sections 
at $Q^2$=2.8 and 4.0 GeV$^2$ \cite{fro99} using a common parameterization 
for the `bare' charge $G_C(Q^2)$ and electric $G_E(Q^2)$ $N\rightarrow\Delta$ quadrupole form factors.  
Near $Q^2=0$, $G_C(0)$ was determined from $G_E(0)$ using the long wavelength 
limit (Siegert's theorem).  The SL curves shown in Fig.~\ref{fig5} describe the $Q^2$ trend of 
the CLAS data reasonably well.  However, the SL model provides a poor fit to
the BATES data \cite{mer01} at $Q^2=0.126$~GeV$^2$ and the SL curve clearly misses the MAMI
$R_{SM}$ point \cite{pos01}.  Those data are better described by the Dubna-Mainz-Taipei dynamical 
model (DMT) \cite{kam01} and a new version of MAID (M2K) \cite{m2000,kam01}, 
(also refitted to the high $Q^2$ data) both of which use different prescriptions for unitarization.  
Although the overall magnitude of the CLAS $R_{EM}$ and $R_{SM}$ measurements is somewhat better described by
DMT, our lowest $Q^2$ point marginally favors the SL prediction. 

The generally successful description of both $R_{EM}$ and $R_{SM}$ 
by the dynamical models strengthens the claim made in  \cite{sat01,kam01} 
that non-resonant meson exchange dominates the $N\rightarrow\Delta(1232)$ quadrupole transition.  This 
has important implications for the interpretation of pure quark model predictions of
photocoupling amplitudes, where pion degrees of freedom are not explicitly treated.
The low $Q^2$ evolution of $E_{1+}$ and $S_{1+}$ is especially important,
since model independent constraints from Siegert's theorem, gauge 
invariance, chiral perturbation theory  \cite{gel99}, and ultimately lattice calculations
can be most accurately applied in this region.

We acknowledge the efforts of the staff of the Accelerator and Physics Divisions at 
Jefferson Lab in their support of this experiment.  This work was supported in part by the U.S. Department 
of Energy, including DOE Contract No. DE-AC05-84ER40150, the National Science Foundation, 
the French Commissariat a l'Energie Atomique, the Italian Istituto Nazionale di Fisica Nucleare, 
and the Korea Research Foundation.

\end{document}